\documentclass[conference]{IEEEtran}
\IEEEoverridecommandlockouts
\usepackage{cite}
\usepackage{amsmath,amssymb,amsfonts}
\usepackage{algorithmic}
\usepackage{graphicx}
\usepackage{textcomp}
\usepackage{xcolor}
\usepackage{tikz}
\usepackage{comment}
\usepackage{subcaption}
\usepackage{float}
\usepackage{lipsum} 
\usepackage{array}
\usepackage{comment}
\usepackage{hyperref}
\usepackage{authblk}

\usepackage[ruled,vlined]{algorithm2e}
\usepackage{filecontents}
\newif\ifcomment
\commenttrue
\ifcomment
\newcommand{\rodrigo}[1]{{\bf \textcolor{blue}{Rodrigo: #1}}}
\newcommand{\shirin}[1]{{\bf \textcolor{purple}{Shirin: #1}}}
\else
\newcommand{\rodrigo}[1]{}
\newcommand{\shirin}[1]{}
\fi

\def\BibTeX{{\rm B\kern-.05em{\sc i\kern-.025em b}\kern-.08em
    T\kern-.1667em\lower.7ex\hbox{E}\kern-.125emX}}
\begin{document}

\title{Audio Attacks and Defenses against AED Systems - A Practical Study}
\author{Rodrigo dos Santos}
\author{Shirin Nilizadeh}
\affil{Department of Computer Science and Engineering, The University of Texas at Arlington}

\maketitle

\begin{abstract}
In this paper, we evaluate deep learning-enabled AED systems against evasion attacks based on adversarial examples. 
We test the robustness of multiple security critical AED tasks, implemented as CNNs classifiers, as well as existing third-party Nest devices, manufactured by Google, which run their own black-box deep learning models. 
Our adversarial examples use \emph{audio perturbations} made of white and background noises. Such disturbances are easy to create, to perform and to reproduce, and can be accessible to a large number of potential attackers, even non-technically savvy ones.

We show that an adversary can focus on audio adversarial inputs to cause AED systems to misclassify, achieving high success rates, even when we use small levels of a given type of noisy disturbance. For instance, on the case of the gunshot sound class, we achieve nearly 100\% success rate when employing as little as 0.05 white noise level. 
We then, seek to improve classifiers' robustness through countermeasures. We employ adversarial training and audio denoising. We show that these countermeasures, when applied to audio input, can be successful, either in isolation or in combination, generating relevant increases of nearly fifty percent in the performance of the classifiers when these are under attack. 
\end{abstract}

\begin{IEEEkeywords}
AED, neural networks, deep learning, spectrograms
\end{IEEEkeywords}

\section{Introduction}

Recently there has been a substantial growth in the use of deep learning for the enhancement of Speech Recognition (SR) and Audio Event Detection (AED) capabilities~\cite{austin2020, eagle2020, abdullah2019practical, choi2005acoustic}.
This generates concerns about the robustness of these classifiers against evasion attacks, where the adversary tries to fool the deep learning model into misclassifying newly seen inputs, thus defeating its purpose. 
Some works have already studied the robustness of deep learning classifiers against different evasion attacks~\cite{carlini2017towards, goodfellow2014explaining}. 
Most of these attacks are focused on either image-based classification tasks~\cite{akhtar2018threat,athalye2017synthesizing,hendrik2017universal,su2019one}, or on speech and speaker recognition applications~\cite{carlini2018audio, kwon2019selective, abdullah2019practical,zhang2017dolphinattack}. Recently, Subramanian et al.~\cite{subramanian2020study} studied the transferability of adversarial attacks in sound event classification, generating adversarial examples based on the Carlini and Wagner attack~\cite{carlini2018audio}. 
This attack keeps perturbing the original audio file until the model misclassifies the audio file. 
However, in a real scenario, it might not be easy to reproduce and inject such perturbation. 
While both Speech Recognition (SR) and AED systems work on audio samples, their goals and algorithms are different. 
Speech recognition works on vocal tracts and structured language, where the units of sound (e.g., phonemes) are  similar. SR systems 
link these  basic  units  of  sounds in  order  to  form  words  and sentences,  which  are  recognizable and meaningful to humans~\cite{bilen2020,cowling2003,Jose2020AccurateDO}. 
However, the AED algorithms cannot look for specific phonetic sequences to identify a specific sound event~\cite{hamid2014}, and because of distinct patterns of different sound events, e.g., dog bark vs. gunshot, a different AED algorithm should be used for every specific sound event. 
Also, for AED, the Signal-To-Noise Ratio (SNR) tends to be low, being even lower when the distance between acoustic source and the microphones performing the audio capture increases~\cite{crocco2016}. 
As such some works argue that developing algorithms for detecting audio events is more challenging~\cite{bilen2020,cowling2003,hamid2014,crocco2016}. 



Given the several critical applications of AED systems, we focus on a home security scenario, where we would deploy an AED system for constantly monitoring the environment for suspicious events, e.g., a window glass is broken, or a gunshot is fired. In our threat model, such AED system is deployed in a physical world, e.g., as part of a home security system, and the adversary, while attempting to cause harm, aims to prevent the AED system to correctly detect and classify the sound events. For this purpose, the adversary generates some noise (e.g., background noise or white noise) which can add perturbations to the audio being captured by the AED system. 
This threat model would demand effort and planning by the attacker, however we consider it to be feasible. For example, in the Mandalay Bay Hotel attack~\cite{nytimesPlan2017}, the shooter had been preparing for the attack for several months. We believe that if this was done in a hotel scenario, such tempering planning can also be reproduced in a less scrutinized, more vulnerable home scenario. For instance, it is not a stretch to envision a scenario where a home burglar could plan for days or weeks in advance on how to deploy attacks against a known audio-based home security system. 

In this work, we consider the AED systems that use Convolutional Neural Networks (\emph{CNNs}), as they are extensively used for implementing AED systems~\cite{Pillos2016ARE, donmoon2017ensembleCNN, matsuoka2020method}. We implemented several different classifiers, where each is capable of detecting one sound event of interest, as well as a multi-classifer that detects multiple events. Our audio classes are diverse and include \emph{gunshots}, \emph{dog bark}, \emph{glassbreak} and \emph{siren}, all of them being representative of sounds that could potentially be considered suspicious if detected in the vicinity of a home. 
x

Through extensive amount of experiments, we evaluated robustness of AED systems against audio adversarial examples, first, by implementing Convolutional Neural Networks (\emph{CNNs}), tailored for AED, and feeding them audio, disturbed by adding different levels of white noise and background noise. Per the categories provided by~\cite{abdullah2020sok}, these would be \emph{over the air signal processing attacks} that generate \emph{noise}. We then performed on-the-field experiments, using real devices manufactured by Google, running their own black-box models, capable of detecting, by the time of the experiments, one type of sound: glass break. Our consolidated results show that AED systems are susceptible against adversarial examples, as the performance of the CNN classifiers as well as of the real devices, in the worst case was degraded by nearly 100\% when tested against the perturbations.

We then implemented some improvements on the classifiers in face of the attacks. The first consisted of adversarial training (adding some disturbed samples to training). The second consisted on a countermeasure based on audio denoising. Adversarial training has been shown to be effective in increasing the performance as well as robustness of image classification tasks~\cite{shorten2019,Wang2019,Song2018}, so we investigate if they also work favorably on audio. Denoising on the other hand, relies on the use of filters for mitigating the audio disturbances. Through more experiments, we could demonstrate the effectiveness of these countermeasure techniques. In particular, our paper has the following contributions:



\begin{itemize}
    \item  
    We evaluate the robustness of deep learning models against adversarial examples that target the audio inputs of the 
    AED tasks, and are easy to generate by the attacker and may not be so noticeable by bystanders;  
    \item We conduct attack field experiments against modern deep learning enabled devices, capable of detecting suspicious events;
    \item We show that deep learning models, deployed in standalone fashion as well as part of real physical devices, are vulnerable against evasion attacks; 
    \item We show that adversarial training and denoising can be used as countermeasures to evasion attacks based on audio. 
\end{itemize}

\section{Threat Model}
While several AED solutions exist~\cite{showen97operationalgunshot,hansheng2103specialsound,Pillos2016ARE,clavel2005eventsdetection,choi2005acoustic,shooter}, we believe that they are still to become truly ubiquitous, possibly powered by massively distributed technologies, such as mobile devices that, thanks to their embedded microphones and sensors, can work as listening nodes. For now, under a smaller range, less distributed, current reality scenario, we choose to focus on home security/ safety audio event detection, given the importance of the topic to a broad audience. 

In this work, we assume that the adversary actively attempts to evade an AED system that aims on detecting suspicious sound events in a home. We assume a black-box scenario, in which the adversary does not have any knowledge about the datasets, algorithms and their parameters. Instead, the adversary uses some sort of gear to generate enough noise disturbances, which will be overlaid to the detectable suspicious sound, being captured together with it, causing the AED system to miss the detection or to misclassify the sound event. 

While AED solutions are still emerging, real physical devices that employ deep learning models for the detection of suspicious events for security purposes are already a reality and have been deployed to homes around the world. 
Some examples of these devices are ones manufactured by major companies, such as the \emph{Echo Dot} and \emph{Echo Show} by Amazon~\cite{Alexa2019}, and \emph{Nest Mini} and \emph{Nest Hub} by Google~\cite{Nest2021a, Nest2021b}. Despite still being limited in terms of detection capabilities, as most of these devices can detect only a few variety of audio events, attempts to create general purpose devices, capable of detecting a wide spectrum of audio events, are known to be in the making, e.g., See-Sound~\cite{see-sound2021}.  

Physical devices that generate audio disturbances on the field are also a reality and are intensively researched and used by military and law enforcement agencies around the world~\cite{popularMechNonAudible2010,digitalTrends2018,nbcNews2019,CNNNewsAudioAttack2017}. For example, gear capable of white noise generation is already largely available to the public~\cite{soundmachines2021}.
Commodity automotive audio gear made of speakers, amplifiers and other components can be easily configured and deployed within, or on top of, almost any commodity vehicle  and could be re-purposed for malicious intents. 
We call these devices as \emph{``Sound Disturbing Devices''} or \emph{SDDs}. 

In out threat model, the audio disturbances used by the adversary do not need to be completely stealthy as even though they would be able to be perceived, it is unlikely they would draw so much attention by individuals near the source of attack, because the audio disturbances could simply be perceived as mere noise, for instance, traffic or music. Even pure white noise would be much less conspicuous then for instance, audible clearly stated voice commands.

As such, our SDDs are limited by research design to generate audible disturbances, and in our threat model, the adversary, when attempting to disrupt the AED system, would generate either audible white or background noises, that would be captured not only by the audio capturing sensors or devices (microphones), but could also be noticeable by people and animals standing close to the source of the disturbance. This noise-infused captured audio becomes our adversarial examples. 

One cannot ignore the apparent heavy planning needed in order to implement such attacks. One cannot also ignore the motivation of adversaries who intend to do harm. For example, the attack that happened in the Mandalay Hotel at Las Vegas~\cite{nytimes2} showcases such motivation, as the attacker spent months smuggling guns and ammunition into his hotel room, and even went to the extent of setting and possibly other sensors in the corridor leading to his room, so he would be better prepared to deal with law enforcement officials when they responded to the emergency situation he was about to set. 
Therefore, it is not a stretch to envision a scenario where a home burglar could plan for days, weeks or even months in advance on how to deploy attacks against an audio-based home security system

\section{Related Work}
\label{related-work}

\textbf{Audio Event Detection Systems.} 
AED systems can collect real-time multimedia data 
and identify audio events.  For example, some surveillance devices identify individual audio events including screams and gunshots~\cite{atrey2006audio, clavel2005events, vu2006audio, geiger2015improving,lieskovska2019acoustic,dufaux2000automatic}. 
Some health monitoring devices detect sounds, such as coughs to identify symptoms of abnormal health conditions~\cite{matos2006detection,larson2011accurate,peng2009healthcare}. 
Some home devices include digital audio applications to 
classify the acoustic events to distinct classes (e.g., a baby cry event, music, news, sports, cartoon and movie)~\cite{matsuoka2020method, vafeiadis2020audio,petridis2010multi, evangelopoulos2009video}. 
Some home security devices also use AED systems~\cite{choi2005acoustic, kawamoto2009system, krstulovic2018audio,atrey2005timeline}. 
Deep Neural Networks (DNNs) have recently been  employed as part of AEDs in order to augment their capabilities. 
Some has explored different feature extraction techniques~\cite{eghbal2016hybrid}, noise reduction techniques~\cite{ragano2020audio,ozer2018noise,mcloughlin2017continuous}, hybrid classifiers~\cite{xu2017unsupervised}, various DNN models~\cite{li2017comparison}, and pyramidal temporal pooling~\cite{zhang2020}. 

\noindent\textbf{Gunshot and Suspicious Sound Detection.} 
Some works, including ones about commercial products~\cite{shooter2020,eagle2020,austin2020,showen97operationalgunshot}, proposed AEDs for gunshot detection. ShotSpotter~\cite{showen97operationalgunshot} and SECURES~\cite{page95securessystem} detect gunshots from distributed sensors deployed to a large coverage area, and by performing signal processing on the acquired data. 
Other works classify \emph{emergency related} sounds through machine learning~\cite{tangkawanit2018developmentOG,hansheng2103specialsound, Pillos2016ARE}
and Neural Networks (NNs)~\cite{tangkawanit2018developmentOG,zhou2017usingdeepconv,khamparia2019soundclassification}.
For the home scenario, glass break detection capabilities are employed 
by Amazon manufactured devices, such as Alexa~\cite{Alexa2019}, and Google devices, such as the \emph{nest hub}~\cite{Nest2021a} and \emph{nest mini}~\cite{Nest2021b}. 

\noindent\textbf{Spectrograms for AEDs.} 
Some works transform audio signals into spectrograms and use them as inputs to the classifiers~\cite{zhou2017usingdeepconv, khamparia2019soundclassification,li2016soundclassificationspectrogram, donmoon2017ensembleCNN, Ghaffarzadegan2017BOSCHRS}. Zhou et al.~\cite{zhou2017usingdeepconv} and Khamparia et al.~\cite{khamparia2019soundclassification} proposed to use a combination of CNN with sequential layers and spectrograms for sound classification.  
Some works~\cite{Lim2017RareSE,cakir2017crnn} use Recurrent NN (RNN) and seek to classify suspicious events. 
In this work, we implemented a modified version of the CNN~\cite{zhou2017usingdeepconv} classification.  
Prior work has shown that NN models trained on images are vulnerable to evasion attacks~\cite{carlini2017towards,erwin2020}.
For this attack the adversary requires to have access to the spectrogram generation portion of AED system, which is not practical and not considered in our threat model.

\noindent\textbf{Adversarial Attacks on Speech Recognition Systems.} 
Speech Recognition Systems are different from AED systems in terms of purpose, implementation, application, vulnerabilities and even how attacks are crafted in order to be effective. SR systems are not the focus of this research
however, 
AED and SR have similarities, and we cannot rule out the possibility that advances on one of these fields in isolation may ultimately lead to advances on the other.   

Recently, a huge body of research focused on the robustness of speech recognition systems against adversarial attacks~\cite{schonherr2018adversarial}. These attacks can be divided into three categories: (1) attacks that generate malicious audio commands that are inaudible to the human ear but are recognized by the audio model~\cite{zhang2017dolphinattack,roy2018inaudible,qin2019imperceptible, chenmetamorph}; (2) attacks that embed malicious commands into piece of legitimate audio~\cite{yan2019feasibility, liADVPulse2020}; and (3) attacks that obfuscate an audio command to such a degree that the casual human observer would think of the audio as mere noise but would be correctly interpreted by the victim audio model~\cite{vaidya2015cocaine, carlini2016hidden, abdullah2019practical}. 
Some works have shown the possibility of attacking Alexa or Google Assistant enabled devices. 
For example,  Li et al.~\cite{junchenl2019AdversarialM} target the local wake-word detection capability of Alexa-enabled devices, employing background music perturbations to disrupt the whole voice assistant functionality.
The features explored in these works are specific to SR systems. For instance, some can be used for keyword-spotting, including sequence of phonemes (and their fragments, called senones), variations in length, speech speed, etc~\cite{Jose2020AccurateDO}. 

In our work, on the other hand, we show that the adversarial examples can be generated by adding noise sounds that are easy to generate by the attacker and are less likely to gain attention as they are common sounds, e.g., music and white noise. In addition, in our lab experiments, we test the attacks on classifiers specific to safety and security domain, and they are obtained using a much larger dataset.  
Moreover, with our field experiments, we show the possibility of such attacks against real AED systems, when attacker only requires to use commodity devices to generate the noise. 
Finally, we examine the effectiveness of adversarial training and propose a denoising technique as a countermeasure to this problem.


\noindent\textbf{Countermeasures against Evasion Attacks.} 
Some work has studied countermeasure techniques for improving the resilience of these systems against adversarial attacks~\cite{roy2018inaudible,carlini2018audio,mao2020watchdog}. Most of these techniques are passive in nature, such as on the case of promoting the detection of an adversarial attack occurrence. 
Active techniques, such as adversarial training exist and can also be found in smaller numbers. For example, Sallo et al.~\cite{sallo2020adversarially}, used six different attacks, all tampering the spectrograms (images) and not the audio files, employing them next against some publicly AED available models. The adversarial training on this case consists of using adversarial spectrograms. 
Also, while adversarial training is a common technique used for increasing the robustness of DL-based classifiers, our proposed denoising technique is unique and novel to this work.

\section{Methodology}

\begin{figure*}[ht]
    \begin{center}
        \includegraphics[width=1\textwidth]{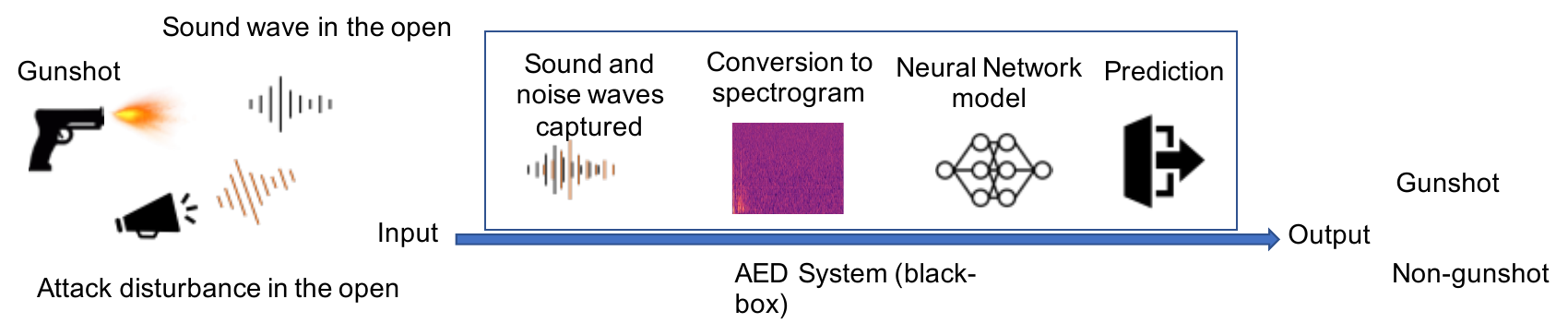}
         \caption{Our framework considers two testing environments: first, AED classifiers implemented based on state-of-the-art algorithms proposed in the literature, and second, third-party AED devices available on the market.}
        \label{fig:1}
    \end{center}
\end{figure*}

As it is shown in Figure~\ref{fig:1}, our framework for evaluating the adversarial robustness of neural network-based AED systems consists of the following modules: (1) data collection, (2) model building, training and testing, (3) generating adversarial examples and testing the models against them, and (4) implementing the countermeasures and testing the classifiers against adversarial examples. 

\subsection{Data Collection}
\label{data}




We focus on AED systems that try to detect suspicious sounds, including \emph{gunshots}, \emph{dog bark}, \emph{glassbreak} and \emph{siren}.  For audio classes that do not contain any sound of interest, we used \emph{pump}, \emph{children playing}, \emph{fan}, \emph{valve} and \emph{music}. These classes are chosen because they can be representative of some of the audio events that could be found near a home scenario, but most likely would be considered to be of benign nature. 

\textbf{Groundtruth dataset creation.} 
We use the following databases: 
    \emph{Detection and Classification of Acoustic Scenes and Events or DCASE dataset}~\cite{DCASE17}: From 2017 and 2018 editions, the DCASE datasets include normalized audio samples with  one single instance of an event of interest happening anywhere inside each audio sample of 30 seconds in length, hence the ``rare'' denomination. Each sample is created artificially, and has background noise made of everyday audio;
      \emph{Urban Sounds Dataset}~\cite{UrbanSound8k}: A database made of everyday sounds found at urban locations. The samples are not normalized and vary quite a bit among themselves;
      \emph{MIMII Dataset}~\cite{UrbanSound8k}: A dataset conceived to aid the investigation and inspection of malfunctioning industrial machines. 
      Some of the sounds in this set can also be found on home scenarios. 
      \emph{Airborne Sound}~\cite{AirborneSound}: An open and free database with audio samples destined to be employed on different sound effects. One such case is that of guns and medieval weapons. The gun part has high quality audio on several different types of guns, recorded from different positions. 
      \emph{Environmental Sounds}~\cite{ESC50}: A dataset of 50 different sound events and over 2,000 samples.
      \emph{Zapsplat}~\cite{Zapsplat}: Over 85,000 professional-grade audio samples as royalties-free music and sound effects.
      \emph{FreeSound}~\cite{FreeSound}: A collaborative database of Creative Commons Licensed sounds.
      \emph{Fesliyan Studios}~\cite{Fesliyan}: A database of royalty-free sounds. 
We call the samples from these datasets that contain audio events of interest (security/ safety related) as \emph{``positive samples''}, and those that do not contain sounds of interest as \emph{``negative samples''}.

\textbf{Data Cleaning and Pre-Processing}. 
We cleaned and preprocessed  the audio, by doing: 
     \noindent (1) \emph{Frequency Normalization}, where the frequencies of all samples are normalized to 22,000 Hertz, to be within the human audible frequency.
    \noindent (2) \emph{Audio Channel Normalization}, where needed, we normalized the number of channels of all samples from stereo to \emph{monaural}, as it is easier to find new samples bearing a single channel.
    \noindent (3) \emph{Audio Length Normalization}, where all samples with less than 3 seconds in length were discarded. 

\begin{figure}[ht]
    \begin{center}
      \includegraphics[width=0.75\columnwidth]{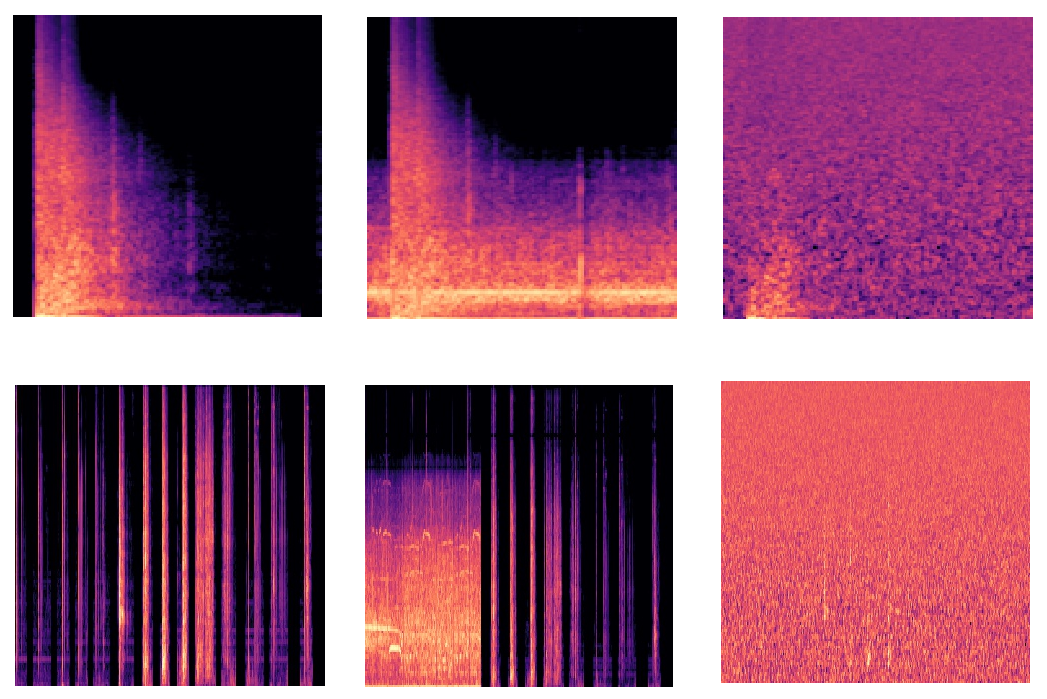}
         \caption{Spectrograms:  
         Left to right, first row: unnoisy gunshot, followed by background noise and white noise disturbed gunshots; second row: unnoisy glass break followed by background and white noise disturbed glass breaks.}
    \label{fig:2}
    \end{center}
\end{figure}

After audio pre-processing, all samples were converted to \emph{spectrograms}, six of these can be seen in Figure~\ref{fig:2}, which are representations of audio in a (usually 2D) graph, that show frequency changes over time for a sound signal, chopping it up and then stacking the spectrum slices, one close to each other. As mentioned in Section~\ref{related-work}, the approach of resorting to spectrograms is a state-of-the-art technique used by AED systems. 
Our spectrogram generating function was implemented through \emph{Librosa}~\cite{Librosa2020}, a python package for music and audio analysis. After the spectrograms are generated, they are vectorized in order to compose a final dataset made of arrays by using \emph{Numpy} library~\cite{Numpy2020}. 
\subsection{Deep Learning based AED Classifiers}

Prior work has shown that CNNs-based AED system perform well~\cite{li2016soundclassificationspectrogram,Lim2017RareSE}. 
Therefore, we focus on evaluating such AED systems. 
Except for one multiclass classifier implementation, we implemented forty-two CNNs as binary classifiers. 
The multiclass classifier was implemented to demonstrate our approach would work under these circumstances and also for performance comparison against the binary classifiers. 

\textbf{Convolutional Neural Networks (CNNs)}. 
CNNs are considered to be the best among learning algorithms in understanding image contents~\cite{khan2019survey}. 
We implemented a CNN model based on the work of Zhou et al.~\cite{zhou2017usingdeepconv}, as it have been successfully used for the purpose of urban sound event detection. Our model is tailored after much experimentation (we have fewer convolutions, more dense layers, different configuration of filters besides a different optimizer). 
The experiments we conducted and bring later prove that our model, after our adaptations, are well tailored for safety audio event detection. This model is consists of: 

(1) \emph{Convolutional layers}: three convolutional blocks with convolutional 2D layers. These layers have 32, 64, 64, 64, 128 and 128 filters (total of 480) of size 3 by 3. Same padding is used on each one of the convolutional blocks.
(2) \emph{Pooling layers}: three 2 by 2 max pooling layers, each coming right after the second convolutional layer of each convolutional block.
(3) \emph{Dense layers}: two dense, aka fully connected layers at the last convolutional block. 
(4) \emph{Activation functions}: ReLU activation is applied after each convolutional layer as well as after the first fully connected layer, while Sigmoid activation is applied only once, after the second fully connected layer. In other words, ReLU is applied to all inner layers, while Sigmoid is applied to the most outer layer.
(5) \emph{Regularization}: applied in the end of each convolutional block as well as after the first fully connected layer, with 25, 50, 50 and 50\% respectively. 
The CNN used binary cross entropy as loss function and RMSprop as optimizer.

\subsection{Third-Party AED Capable Devices}

As a second testing environment, we evaluate some third-party AED capable devices. We chose devices that are readily available on the market. 
Given the well-known involvement of Google with Deep Learning (e.g.,  creation and release of TensorFlow), and the fact that Google AI-enabled devices, including Nest devices are already widely used in day-to-day life~\cite{policyadvice2021}, we test the following devices:  
 \textbf{Nest Mini}: From the large variety of Nest devices available, we started by choosing the most basic device possible, the \emph{Nest mini}~\cite{Nest2021a}. The Nest mini device, currently in its second generation~\cite{analyticsindiamag2019}, and already includes a machine learning chip capable of implementing advanced techniques such as natural language processing and speech recognition. 
Yet another advantage of these devices is the fact that they can work in pairs, in theory augmenting their detection capabilities. 
 \textbf{Nest Hub}: 
We also use the \emph{Nest hub}~\cite{Nest2021b} device, which 
offers all Nest mini capabilities and a display~\cite{pocketlint2021}. Nest hub can be an attractive device to consumers who want to start their own smart home implementation with some simplicity, but want something more refined and capable than the simple Nest mini.  

\subsection{Evasion Attacks}
An adversarial example is defined as a sample of input data which has been modified in a way that is intended to cause a machine learning algorithm to misclassify it~\cite{kurakin2016adversarial}. 
We implement two variants of evasion attacks based on two forms of audio noise, namely \emph{background} and \emph{white} noise. 
Both are easy and fast to generate, able to be crafted practically on the spot of the attack. 

Every practical application that deals with audio signals also deals with the issue of noise. As stated by Prasadh et al.~\cite{prasadh2017efficiency}, ``natural audio signals are never available in pure, noiseless form.'' As such, even under ideal conditions, natural noise may be present in the audio being in use. We use \emph{White noise}, which as pointed out by~\cite{edmonds2006abstraction}, happens when ``each audible frequency is equally loud,'' meaning no sound feature, ``shape or form can be distinguished.''
While other types of noise, such as \emph{Gaussian noise}, and \emph{Pink noise}, can be used by an adversary, we chose \textbf{white noise}, since this noisy variant is widely adopted by different research across~\cite{dahlan2018,vashuki2012automaticnoise, montillet2013extractingwhite}. 
We also considered \textbf{background noise}. This variant is represented by all sorts of noise occurring during the normal course of business, and that may overlay to any sound of interest. Examples of such noise would be that of people talking, active vehicle traffic, music playing, etc.  

Note that in our tests, both forms of noise are added to the audio samples, and only then other subsequent processing will occur, including the required spectrogram generation, ensuring the practicality of the attack. 
The same holds true for the third-party equipment tests, as the disturbances are added when these devices are actively listening for glass break sounds, being introduced to the environment through loudspeakers.
The pseudo-code in Algorithm~\ref{alg1} and Algorithm~\ref{alg2} shows the mechanism for the addition of background and white noises to a given audio sample. 

In Algorithm~\ref{alg1}, two separate files are retrieved, one with the sound of interest, and one with the background noise. Such background noise is added to the sound of interest without any modification of variability, except for the adjustment factor, that simply controls the amplitude (or loudness) of the noise.  In Algorithm~\ref{alg2} white noise is added to the original audio sample, while again configuring it with the amount of desired noise (through the adjustment factor or amplitude control), however, unlike in the case of background noise, where a separate file is needed, the white noise is derived from the highest amplitude already present in the sound sample being disturbed. 

We used different thresholds for this adjustment factor, ranging from 0 (no white noise) and 1 (100 percent noise), thus generating multiple thresholds along this interval, particularly, \emph{0.0001, 0.0005, 0.001, 0.005, 0.01,  0.05, 0.1, 0.2, 0.3, 0.4,} and \emph{0.5}.  
The initial and final thresholds were chosen based on trial and error, where we were able to find, either disturbance values that were mild in terms of perceptibly while still being able to fool the classifiers, or were much more pronounced, while also causing severe disruptions to classifier performance. The remaining thresholds were chosen in the interval between these initial and final values.

\begin{algorithm}[t]
\SetAlgoLined
\small{
\KwResult{Perturbed audio sample}
 initialization\;
 \For{number of audio files in the test set}{
  sample = load audio file as an array\;
  noise = load audio file as an array\;
  adjusted noise = noise + adjustment factor\;
  perturbed sample = sample + adjusted noise\;
  save perturbed sample\;
  }
 }
 \caption{Background Noise Generation}
 \label{alg1}
\end{algorithm}

\begin{algorithm}[t]
\SetAlgoLined
\small{
\KwResult{Perturbed audio sample}
 initialization\;
 \For{number of audio files in the test set}{
  sample = load audio file as an array\;
  noise = adjustment factor * max element of the array\;
  perturbed sample = sample + noise * normal~distribution\;
  save perturbed sample\;
 }
 }
 \caption{White Noise Generation}
 \label{alg2}
\end{algorithm}

\subsection{Countermeasures}
We investigated techniques for increasing the robustness of these systems against adversarial examples. We implemented and evaluated two mechanisms: (i) \emph{adversarial training} and (ii) \emph{denoising}. 

\textbf{Adversarial Training}. 
This is a popular technique applied by several researchers~\cite{Wang2019,Song2018}. It  consists of introducing some adversarial examples into the training set, thus leading to  increased resilience against adversarial attacks through learning directly from adversarial examples. 
While adversarial learning has been mostly been used for image classification tasks, we examine its effectiveness for AED systems. 

\textbf{Audio Denoising}. 
Our second countermeasure uses audio denoising techniques to remove or at least mitigate the noise previously introduced to the disturbed samples. Other works have used filters to perform audio denoising, thus leading to improvement in classifier's performance. Some works~\cite{kiapuchinski2012,Hodgson2010,Audacity2020} used some variation of a technique called \emph{Spectral Noise Gating}~\cite{Hodgson2010}. Such work consists of performing the reduction of a signal found to be below a given threshold (the noise), and an important point about it was brought up by Kiapuchinski et al.~\cite{kiapuchinski2012}, consisting of its requirement to have a noise profile (extracted from the known noise), from which a smoothing factor will be derived and applied to the signal that requires denoising (the whole sound).
 We show a high-level pseudo-code of such denoising function as Algorithm~\ref{denoise}. 

\begin{algorithm}[t]
\SetAlgoLined
\small{
\KwResult{Perturbed audio sample}
 initialization\;
 \For{number of audio files in the test set}{
  sample = load audio file as an array\;
  noise = load audio file as an array\;
  sample profile = calculate statistics specific to sample\;
  noise profile = calculate statistics specific to noise\;
  \uIf{sample profile < noise profile}{
    apply smoothing filters\;
    save denoised sample\;
    }
 }
 }
 \caption{Denoising Algorithm}
 \label{denoise}
\end{algorithm}


Our custom denoising spectral gating function is based specifically on the noise reduction function employed by Audacity~\cite{Audacity2021}, the open-source digital audio editor and recording application software, and rewritten in python code~\cite{Sainburg2018}. 
The original function requires two input sounds in order to perform denoising (one with noise, with the audio to be denoised). However, in practice, the defender does not have any knowledge about the noise function used by the adversary. 
We overcame this challenge by using the same ``whole sound'' as both audio and noise profiles donor.
Our implementation also brings additional changes, for instance, ones related to 
\emph{frequency channels}, \emph{fourier transform frames}, \emph{window and hop length}, and \emph{time and frequency smoothing filter setup.}
These changes were performed after much experimentation. 
We believe these changes, even though they are not definitive (as they can be further improved in the next iterations of this research) are fundamentally in the right direction. The modified algorithm allows us to reduce the noise fingerprint on each frequency spectrum of the audio, while at the same time representing a better tailored approach for our AED domain problem at hand. 



\section{Experimenterimental Setup}
\label{model_testing} 
We proceeded with the design and preparations of several experiments needed to test the AED capabilities and their robustness against adversarial examples. 
In the case of our own implementation, besides crafting the several training and testing sets, we actually train our CNN models. In the case of the third party devices, we set them up in a controlled environment, reproduce glass break sounds on the field, and also attack the devices with the intention of crippling their detection capabilities. 

Our experiments were largely binary, except for one multiclass instance and we used \emph{bark, glassbreak, gun} and \emph{siren} as positive classes, and \emph{pump, children playing, fan, valve} and \emph{music} as negative classes. 
Both the training and test sets always had the two participating classes in a balanced fashion. In other words, we always made sure to have the same amount of samples per class in each experiment. 
A summary of our experiments follows next. 

\subsection{Experiment 1 - Baseline}

These experiments involved only pure audio as inputs, meaning no disturbances were introduced as part of the testing procedure. These  tests set the baseline model performance against which we compare most of the upcoming experiments.

 \begin{itemize}
 \item 
\textbf{Experiment 1a - Binary CNN Classifiers}: 
We trained 4 binary models, each with 1000 positive samples and 1000 negative samples. The positive samples in each model belong to one of the categories of sounds, i.e., \emph{bark, glassbreak, gun} and \emph{siren}. 
The negative portion of the training set was kept unaltered throughout the 4 experiments, and was made of a combination of 200 samples of each one of the five different negative classes previously presented. The respective test sets were made of 300 samples, 150 positive, and 150 negative. 

 \item 
\textbf{Experiment 1b - Multiclass CNN Classifier}: 
This experiment involved a multiclass version of the CNN algorithm, including now all 4 positive classes at once. Our goal is to investigate if multiclass classifiers provide different results or show different behavior compared to binary classifiers, even though currently readily available AED systems are dedicated to detect one or two audio classes only.  In this experiment, the training sets were made of the 4,000 positive samples used in Exp 1a, with no negative classes.

 \end{itemize}

\subsection{Experiment 2 - Third-Party Devices}
\begin{figure*}[ht]
    \begin{center}
        \includegraphics[width=0.8\textwidth]{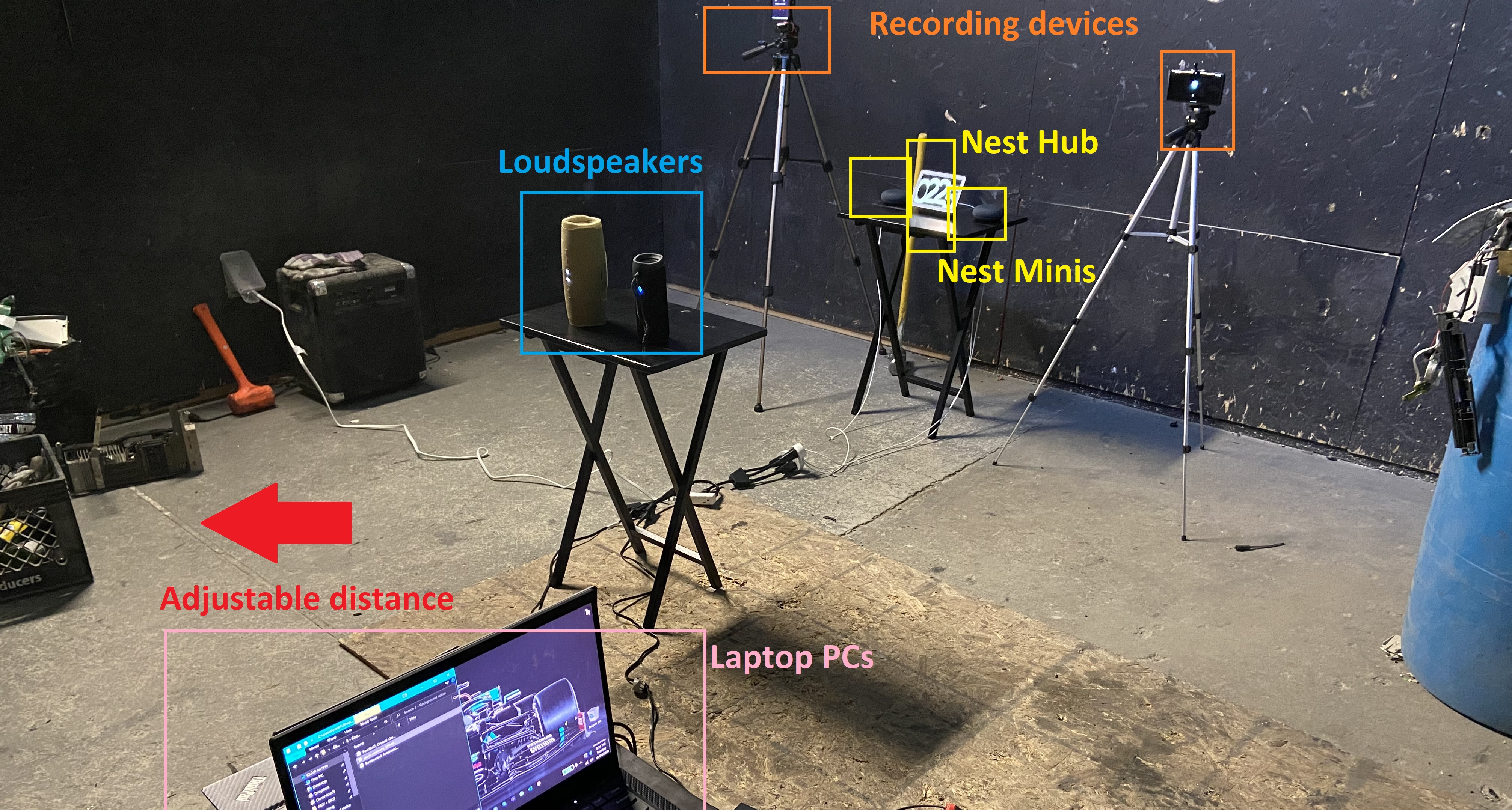}
        \caption{Testing third-party devices with AED capabilities against adversarial examples. Note that the glass items to be broken are not captured in the picture.}
        \label{fig:3}
    \end{center}
\end{figure*}  

Figure~\ref{fig:3} shows our experimental setup for testing third-party devices in practice. 
These experiments involve the use of Google nest hub and minis, set as a representation of an implementation of an audio monitoring home security system. All experiments were conducted at the city of Break Stuff~\cite{breakstuff}, in the city of San Jose, California. From 2a to 2c,  we used a single \emph{nest hub} and two \emph{nest mini} devices, initially working in isolation from each other, later working together, in order to detect glass break sounds. As attacking devices, we used two easy to carry loudspeakers, namely \emph{Charge 4} and \emph{Flip 4}, both manufactured by \emph{JBL}, positioned at 2 and 4 meters from the nest devices. 

For real glass break sounds, we broke previously purchased glass items, such as bottles, cups and plates. To record the whole procedure, but also to allow later reuse during experiment 2d of the real glass break sounds being captured, we employed two Android devices, namely S10+ and S20 Ultra, working as audio recorders, positioned at negligible distance from the nest devices. To establish signal-to-noise ratio readings, the room where the experiments were conducted was recorded when being free of any experiment-related sound, measuring 60 decibels by then.

We then connected each portable loudspeaker via bluetooth wireless protocol to an identical Lenovo X1 Carbon laptop computer, one holding the white noise disturbances, the other holding the background noise disturbances. The sound volume on both computers was set to 100 percent while the loudspeakers had their volume set at 50 percent. We played the disturbances and remeasured the new signal-to-noise ratios, now measuring 70 decibels when the background noise was played and 75 decibels when the white noise was played.
In summary, we ran the following experiments: 

 \begin{itemize}
 \item 
\textbf{Experiment 2a - Digital Pure Audio Inputs}: 3rd-party devices exposed to digital glass break sounds, without any disturbance being played through the loudspeakers.
 \item 

\textbf{Experiment 2b - Real Pure Audio Inputs}: 3rd-party devices exposed to real glass break sounds, without any disturbance being played through the loudspeakers.
 \item 

\textbf{Experiment 2c - Background Noise Disturbed Inputs}: 3rd-party devices exposed to real glass break sounds, with background noise disturbance being played through a loudspeaker. 

 \item 
\textbf{Experiment 2d - White Noise Disturbed Inputs}: 3rd-party devices exposed to real glass break sounds, now with white noise disturbance being played through a loudspeaker. 

 \item 
\textbf{Experiment 2e - Binary CNN Classifier and Pure Glass Break Recordings}: The CNN classifiers, now being fed, during test phase, with glass break sounds recorded during experiments 2a, 2b and 2c, by the S10+ and S20 Ultra devices.  

\end{itemize}

\subsection{Experiment3 - Audio Adversarial Examples}
Going forward we focus on two positive classes, namely gunshot and glass break. 
We test the same two respective, previously trained gunshot and glass break classifiers, against increasing levels of background and white noises. For the background noise, we used Pydub python library to digitally add two different background noises, namely \emph{car traffic} and \emph{people talking}, to the testset samples to be fed to the models. To emphasize, these background noises are not related to the \emph{negative} classes that used to train and test the classifiers. Therefore, if the models misclassify the adversarial samples generated via background noise, it is not due to existence of similar samples in the negative class. 

We kept the signal-to-noise ratio at 10 decibels, similarly to the on-the-field experiments on third-party devices. 
We used Numpy library to digitally generate white noise disturbances, and we used Librosa and SoundFile libraries to add the disturbances to the testset samples. By doing so we crafted eleven different testsets, each having 100\% of their samples overlaid with 0.0001, 0.0005, 0.001, 0.005, 0.01, 0.05, 0.1, 0.2, 0.3, 0.4 and 0.5 white noise levels.

 \begin{itemize}
 \item 
\textbf{Experiment 3a - Glass break Classifier and Background Noise Infused Audio Inputs}: Glass break classifier from Experiment 1, tested against three different testsets, having 25\%, 50\% and 100\% of their samples infused with background noise. 

\item 
\textbf{Experiment 3b - Gunshot Classifier and Background Noise Infused Audio Inputs}: Gunshot classifier from Experiment 1, tested against three different testsets, having 25\%, 50\% and 100\% of their samples infused with background noise.

 \item 
\textbf{Experiment 3c - Glass break Classifier and White Noise Infused Audio Inputs}: Glass break classifier from Experiment 1, tested against the eleven different white noise infused testsets.

 \item 
\textbf{Experiment 3d - Gunshot Classifier and White Noise Infused Audio Inputs}: Gunshot classifier from Experiment 1, tested against the eleven different white noise infused testsets.
 \end{itemize}

\subsection{Experiment4 - Background Noise for Adversarial Training}
We test the effectiveness of adversarial training as a countermeasure against evasion attacks, when background noise infused samples are added to training sets. 

 \begin{itemize}
 \item 
\textbf{Experiment 4a - Glass Break with Background Noise}: 
from Experiment 3a, we use its 100 percent background noise infused glass break test set, and we modify its train set, now turning 25, 50 and 100 percent of its samples, into adversarial examples by infusing them with background noise.
 \item 
\textbf{Experiment 4b - Glass break Oversampled Background Noise}: From experiment 3a, we use the same 100 percent background noise infused glass break test set, and we modify its train set, as we join Experiment 3a and Experiment 5a train sets. The resulting train set is made, as such, of half pure samples and half disturbed samples.

 \item 
\textbf{Experiment 4c - Gunshot with Background Noise}: from Experiment 3b, we use its 100 percent background noise infused gunshot test set, and we modify its train set, now turning 25, 50 and 100 percent of its samples, into adversarial examples by infusing them with background noise.

 \item 
\textbf{Experiment 4d - Gunshot Oversampled Background Noise }: From experiment 3b, we use the same 100 percent background noise infused gunshot test set, and we modify its train set, as we join experiments 3b and 5b train sets. The resulting train set is made, as such, of half pure samples and half disturbed samples.
 \end{itemize}

\subsection{Experiment5 - White Noise Adversarial Training}

We test the effectiveness of adversarial training based as a countermeasure to evasion attacks, when white noise infused samples are added to the train sets. 

 \begin{itemize}
 \item 
\textbf{Experiment 5a - Glass break with White Noise}: We use all the the eleven glass break test sets from Experiment 3c, and we modify the glass break train set from Experiment 1a, adding to it, proportionally, ten out of the eleven white noise levels previously used (0.0005 to 0.5). As such, every white noise level had one hundred samples included in 6a train set.
 \item 

\textbf{Experiment 5b - Gunshot with White Noise}: We use all the the eleven gunshot test sets from experiment 3d, and we modify the gunshot train set from Experiment 1a, adding to it, proportionally, ten out of the eleven white noise levels previously used (0.0005 to 0.5). As such, every white noise level had one hundred samples included in 6b train set.
 \end{itemize}

\subsection{Experiment6 - Denoising Background Noise} 
We test our experimental denoising algorithm which is based on Spectral Gating. 

 \begin{itemize}
 \item 
\textbf{Experiment 6a - Glass break Testsets}: From Experiment 1a, we take the original, free-of-noise glass break train set, and from Experiment 3a we take the 100\% background noise infused test set, proceeding next to denoise it, thus generating a denoised glass break test set.

 \item 
\textbf{Experiment 6b - Gunshot Testsets}: From experiment 1a, we take the original, free-of-noise gunshot train set, and from Experiment 3b we take the 100\% background noise infused test set, denoise it, thus generating a denoised gunshot test set.
 \end{itemize}

\subsection{Experiment7 - Denoising White Noise}
 \begin{itemize}
 \item 
\textbf{Experiment 7a - Glass break Testsets}: From Experiment 1a, we take the original, free-of-noise glass break train set, and from Experiment 3c, we take all eleven white noise infused test sets, denoise them, thus generating denoised glass break test sets. 

 \item 
\textbf{Experiment 7b - Gunshot Testsets}: From Experiment 1a, we take the original, free-of-noise gunshot train set, and from Experiment 3d, we take all eleven white noise infused test sets, denoise them, thus generating denoised gunshot test sets. 
 \end{itemize}

\section{Results}

\subsection{Baseline Results with Pure Sounds}

\begin{table*}[h!]
\centering
\caption{Baseline Tests with CNN-based AED System}
\resizebox{0.65\textwidth}{!}{%
\begin{tabular}{|c|l|c|c|c|c|c|c|}
\hline
\textbf{AED System}    & \textbf{Exp. Id}               & \textbf{Train Samples)}   & \textbf{Test Samples}    & \textbf{Ac}    & \textbf{Pr}   & \textbf{Rc}   & \textbf{F1}  \\ \hline \hline
 & 1a - Bark - digital   & 2000  & 300  & 0.96 & 0.96        & 0.96         & 0.96\\ 
                      & 1a - Glass break  - digital    & 2000                      & 300                      & 0.99         & 0.99        & 0.99        & 0.99\\ 
Custom CNN             & 1a - Gun  - digital            & 2000                      & 300                      & 0.99           & 0.99        & 0.99          & 0.99\\
                      & 1a - Siren - digital           & 2000                      & 300                      & 0.94         & 0.94         & 0.94         & 0.94\\ 
                      & 1b - Multiclass - digital      & 4000                      & 600                      & 0.93         & 0.93        & 0.93        & 0.93\\ \hline
Custom CNN             & 2e - Glass break - real        & 2000                      & 150                      & 1              & 1             & 1             & 1\\ \hline
\end{tabular}
}
\label{table:table1}
\end{table*}

\begin{table*}[h!]
\centering
\caption{Tests with Third-Party AED-Capable Systems}
\resizebox{0.7\textwidth}{!}{%
\begin{tabular}{|c|l|c|c|c|c|c|c|}
\hline

\textbf{AED System}    & \textbf{Exp. Id}                    & \textbf{Attempts}      & \textbf{Detected}     & \textbf{Missed}    & \textbf{Detection Success Rate}     \\ \hline \hline
                       & 2a - Glass break - digital          & 15                     & 0                     & 15                 & 0\%  \\ 
3rd Party              & 2b - Glass break (unnoisy) - real   & 18                     & 6                     & 12                 & 33\% \\ 
Devices                & 2c - Glass break \& BN - real        & 18                     & 2                     & 16                 & 11\% \\ 
                       & 2d - Glass break \& BN - real        & 12                     & 1                     & 11                 & 8.3\% \\ \hline
\end{tabular}
}
\label{table:table2}
\end{table*}

\textbf{The Performance of CNN Classifiers for AED}. 
Table~\ref{table:table1} shows that the base classifiers, trained only on noise-free samples, present great performance. The four binary classifiers, namely \emph{dog barking}, \emph{glass breaking}, \emph{gunshots} and \emph{siren}, all perform above 94\% accuracy, while the multiclass classifier that includes all these same classes at once, also performs well, having an accuracy of close to 93\%. 
Therefore, the multiclass classifier is on par with the binary classifiers. 

Given the satisfactory baseline performance presented, and also taking into account the large number of experiments, going forward, we narrow down our positive classes to the two best performing ones, namely (glass break and gunshot). Also, since the binary and multiclass classifiers show roughly equivalent performance, going forward we solely conduct binary experiments. Finally, it is important to consider that the third-party devices to be tested are capable of detecting \emph{glass break} sounds, which is a major incentive for us to keep this class as part of the upcoming experiments.         

\textbf{The Performance of Third Party Devices}. 
We started the test of third-party devices, namely Nest mini and Nest hub, without knowing what to expect. The first tests involved checking if said devices, isolated from each other or working in combination, would get their detection capabilities triggered by digital samples (non-real glass break). As such, using the laptop computers and the loudspeakers, we reproduced fifteen glass break sounds, five of them for a single Nest mini, five of them for two Nest minis, and five of them for the two Nest minis plus the Nest hub.

Table~\ref{table:table2} shows that none of the fifteen digital samples triggered any of the Nest devices. This indicates that these devices are well calibrated for detecting real sounds only. Therefore, we proceeded next to break real glass break devices, seeking to assess how well the devices perform in the practice. For this experiment in particular, we broke a total of 48 glass items, eighteen of them under unnoisy conditions, further eighteen when a loudspeaker was playing background noise, and finally, twelve when a loudspeaker was playing white noise. Out of forty-eight, twenty-four breakages happened at one meter (39.3 inches) away from the Nest devices and the other twenty-four happened at 2 meters (78.7 inches) away.

As it can be seen from Experiments 2b, 2c and 2d in Table~\ref{table:table2}, even under unnoisy conditions, the Nest devices perform poorly, with a detection rate of about 33\%, which only gets worse when disturbances are introduced to the environment. Particularly, the background noise is able to reduce detection rates by 22\% while white noise reduces them by 25\%. This is concerning as families may trust their security and safety to these devices to some extent. 
We also experimented from different distances, however, we could not verify any distinct performance change for different setups.  

Finally, as part of experiment 2e, we use a subset of the real glass break sounds recorded by the S10 and S20 devices (75 in total), and use them to test the previously in-house trained glass break CNN classifier. Under these circumstances, the CNN model had an even higher detection accuracy, now of one hundred percent.

\subsection{Evasion Attacks against CNN Classifiers}  
This section is dedicated to the experiments involving adversarial examples for both attack and defense purposes. 

\begin{table}[h!]
\centering
\caption{Adversarial Attack Tests with Adversarial Examples Against Custom CNN AED System. The size of training and testing sets in all the models are 2000 and 300, respectively.}
\resizebox{0.4\textwidth}{!}{%
\begin{tabular}{|c|l|c|c|c|c|}
\hline
\textbf{AED System}    & \textbf{Exp.}  & \textbf{Ac}    & \textbf{Pr}   & \textbf{Rc}   & \textbf{F1}  \\ \hline \hline

Glass break            & Baseline (1a)                        & 0.99         & 0.99        & 0.99        & 0.99\\ \hline

Gunshot             & Baseline (1a)                                & 0.99           & 0.99        & 0.99          & 0.99\\\hline

Glass break            & 3a - 25\% BN                              & 0.88         & 0.90         & 0.88        & 0.87\\ 
-                      & 3a - 50\% BN                               & 0.76         & 0.84        & 0.76        & 0.75\\ 
(digital)              & 3a - 100\% BN                              & 0.71         & 0.82        & 0.71        & 0.69\\ \hline
Gunshot                & 3b - 25\% BN                               & 0.96         & 0.93        & 0.96        & 0.96\\ 
-                      & 3b - 50\% BN                               & 0.94         & 0.95        & 0.94        & 0.94\\ 
(digital)              & 3b - 100\% BN                              & 0.92         & 0.93        & 0.92        & 0.92\\ \hline
                       & 3c - 0.0001 WN                             & 0.99         & 0.99        & 0.99        & 0.98\\ 
                       & 3c - 0.0005 WN                             & 0.95         & 0.96        & 0.95        & 0.95\\ 
                       & 3c - 0.001 WN                              & 0.95         & 0.95        & 0.95        & 0.94\\ 
                       & 3c - 0.005 WN                              & 0.98         & 0.98        & 0.98        & 0.98\\ 
Glass break            & 3c - 0.01 WN                               & 0.99         & 0.99        & 0.99        & 0.98\\ 
-                      & 3c - 0.05 WN                               & 0.93         & 0.93        & 0.93        & 0.93\\ 
(digital)              & 3c - 0.1 WN                                & 0.81         & 0.86        & 0.81        & 0.80\\ 
& 3c - 0.2 WN                                & 0.81         & 0.86        & 0.81        & 0.80\\ 
                       & 3c - 0.3 WN                                & 0.66         & 0.8         & 0.66        & 0.62\\ 
                       & 3c - 0.4 WN                                & 0.65         & 0.79        & 0.65        & 0.60\\ 
                       & 3c - 0.5 WN                                & 0.61         & 0.78        & 0.61        & 0.54\\ \hline
                       & 3d - 0.0001 WN                             & 0.98         & 0.98        & 0.98        & 0.98\\ 
                       & 3d - 0.0005 WN                             & 0.85         & 0.88        & 0.85        & 0.84\\ 
                       & 3d - 0.001 WN                              & 0.9            & 0.92        & 0.9           & 0.9\\ 
                       & 3d - 0.005 WN                              & 0.66           & 0.8         & 0.66          & 0.66\\ 
Gunshot                & 3d - 0.01 WN                               & 0.63         & 0.79        & 0.79        & 0.57\\ 
-                      & 3d - 0.05 WN                               & 0.59         & 0.77        & 0.59        & 0.5\\ 
(digital)              & 3d - 0.1 WN                                & 0.58           & 0.77        & 0.58          & 0.49\\ 
                       & 3d - 0.2 WN                                & 0.55         & 0.76        & 0.55        & 0.43\\ 
                       & 3d - 0.3 WN                                & 0.54         & 0.76        & 0.54        & 0.41\\ 
                       & 3d - 0.4 WN                                & 0.52         & 0.76        & 0.52        & 0.38\\ 
                       & 3d - 0.5 WN                                & 0.5         & 0.75        & 0.5        & 0.34\\ \hline
\end{tabular}
}

\label{table:table3}
\end{table}

\begin{table}[h!]
\centering
\caption{Adversarial Training Defensive Tests. The size of training and testing sets in all the models are 2000 and 300, respectively.}
\resizebox{0.42\textwidth}{!}{%
\begin{tabular}{|c|l|c|c|c|c|}
\hline
\textbf{AED System}    & \textbf{Exp.}                 &  \textbf{Ac}    & \textbf{Pr}   & \textbf{Rc}   & \textbf{F1}  \\ \hline \hline
Glass break  & 4a - 100\% BN                  & 1              & 1             & 1             & 1\\  \hline
Gunshot      & 4b - 100\% BN                                       & 1              & 1             & 1             & 1\\ \hline
                       & 5a - 0.0001 WN                                & 0.99           & 0.99         & 0.99        & 0.99\\ 
                       & 5a - 0.0005 WN                                      & 0.99           & 0.99        & 0.99          & 0.99\\ 
                       & 5a - 0.001 WN                                       & 0.99           & 0.99        & 0.99        & 0.99\\ 
                       & 5a - 0.005 WN                                       & 1              & 1             & 1             & 1\\ 
Glass break            & 5a - 0.01 WN                                        & 1              & 1             & 1             & 1\\ 
                     & 5a - 0.05 WN                                        & 1              & 1             & 1             & 1\\ 
             & 5a - 0.1 WN                                         & 1              & 1             & 1             & 1\\ 
                       & 5a - 0.2 WN                                         & 1              & 1             & 1             & 1\\ 
                       & 5a - 0.3 WN                                         & 1              & 1             & 1             & 1\\ 
                       & 5a - 0.4 WN                                         & 1              & 1             & 1             & 1\\ 
                       & 5a - 0.5 WN                                         & 1              & 1             & 1             & 1\\ \hline
                       & 5b - 0.0001 WN                                      & 0.98         & 0.98        & 0.98        & 0.98\\ 
                       & 5b - 0.0005 WN                                      & 0.98           & 0.98          & 0.98          & 0.98\\ 
                       & 5b - 0.001 WN                                       & 0.99        & 0.99       & 0.99        & 0.99\\ 
                       & 5b - 0.005 WN                                       & 0.99         & 0.99        & 0.99        & 0.99\\ 
Gunshot                & 5b - 0.01 WN                                        & 0.99         & 0.99        & 0.99        & 0.99\\ 
-                      & 5b - 0.05 WN                                        & 0.997         & 0.997        & 0.997        & 0.997\\ 
              & 5b - 0.1 WN                                         & 0.997         & 0.997        & 0.997        & 0.997\\ 
                       & 5b - 0.2 WN                                         & 0.997         & 0.997        & 0.997        & 0.997\\ 
                       & 5b - 0.3 WN                                         & 0.997         & 0.997        & 0.997        & 0.997\\ 
                       & 5b - 0.4 WN                                         & 0.997         & 0.997        & 0.997        & 0.997\\ 
                       & 5b - 0.5 WN                                         & 0.997         & 0.997        & 0.997        & 0.997\\ \hline

\end{tabular}
}
\label{table:adv-training}
\end{table}

\textbf{Generating Adversarial Examples with Background Noise}. 
Experiments 3a and 3b are based on background noise as an attacking mechanism. As such, from Experiment 1a, we reused the \emph{glass break} and \emph{gunshot} baseline classifiers as well the test sets, except that we modify these sets by progressively increasing the number of samples within them that are infused with background noise. 
The results 
can be seen in Table~\ref{table:table3}, which shows the effectiveness of the background noise disturbances, as they increasingly affect classifier's performance. 
The results produced are not even, since the glass break classifier performs worse to the disturbances, presenting an accuracy drop of up to 28\% when 100\% of the test set is infused with background noise. Note that the noise is added to only the samples in the positive class, e.g., gun, glass break. 
In contrast, the gunshot classifier has its performance dropping by around 7\%. 

Different performance drops on different classes due to background noise was expected, as the effectiveness of these disturbances will be affected by several factors, for instance, how loud the sound of interest is to begin with.  We believe this to be the primary reason for the difference on these particular experiments involving gunshot and glass break (the first being much louder and distinct than the second). 

\textbf{Generating Adversarial Examples with White Noise}. 
We adopt the same approach adopted during previous Experiments 3a and 3b, and as such we reuse the glass break and gunshot baseline classifiers as well their test sets, but now we infuse all test samples with progressively higher white noise levels, ranging from 0.0001 to 0.5. The whole list of white noise levels as well as the experiment results are disclosed in Table~\ref{table:table3}. Based on these results, the gunshot sounds prove to be more susceptible to the white noise disturbances than glass break, presenting sharp accuracy drops of over 40\%. 
We observe that white noise-infused adversarial examples significantly decrease the performance of both the gunshot and glassbreak classifiers.

\begin{table}[h!]
\centering
\caption{Denoising Defensive Tests}
\resizebox{0.4\textwidth}{!}{%
\begin{tabular}{|c|l|c|c|c|c|}
\hline
\textbf{AED System}    & \textbf{Exp. Id.}    & \textbf{Ac}    & \textbf{Pr}   & \textbf{Rc}   & \textbf{F1}  \\ \hline \hline
Glass break               & 6a - 100\% BN                                       & 0.74           & 0.83        & 0.74          & 0.72\\ \hline
Gunshot               & 6b - 100\% BN                                       & 0.94         & 0.95        & 0.94        & 0.94\\ \hline  
                       & 7a - 0.0001 WN                                      & 0.99         & 0.99        & 0.99        & 0.99\\ 
                       & 7a - 0.0005 WN                                      & 0.97         & 0.97        & 0.98        & 0.98\\ 
                       & 7a - 0.001 WN                                       & 0.95         & 0.96        & 0.95        & 0.95\\ 
                       & 7a - 0.005 WN                                       & 0.97           & 0.97        & 0.97          & 0.97\\ 
Glass break            & 7a - 0.01 WN                                        & 0.97         & 0.97        & 0.97        & 0.97\\ 
                     & 7a - 0.05 WN                                        & 0.57           & 0.7           & 0.57          & 0.49\\ 
              & 7a - 0.1 WN                                         & 0.96           & 0.98          & 0.96          & 0.96\\ 
                       & 7a - 0.2 WN                                         & 0.98         & 0.98        & 0.98        & 0.98\\ 
                       & 7a - 0.3 WN                                         & 0.98           & 0.98        & 0.98          & 0.98\\ 
                       & 7a - 0.4 WN                                         & 0.98           & 0.98        & 0.98          & 0.98\\ 
                       & 7a - 0.5 WN                                         & 0.98           & 0.98        & 0.98          & 0.98\\  \hline
                       & 7b - 0.0001 WN                                      & 0.98         & 0.98        & 0.98        & 0.98\\ 
                       & 7b - 0.0005 WN                                      & 0.85         & 0.88        & 0.85        & 0.84\\ 
                       & 7b - 0.001 WN                                       & 0.91           & 0.92        & 0.91        & 0.91\\ 
                       & 7b - 0.005 WN                                       & 0.68          & 0.82        & 0.71        & 0.69\\ 
Gunshot                & 7b - 0.01 WN                                        & 0.66         & 0.66        & 0.66        & 0.66\\ 
                     & 7b - 0.05 WN                                        & 0.62           & 0.79        & 0.63        & 0.57\\ 
             & 7b - 0.1 WN                                         & 0.60         & 0.60        & 0.60        & 0.53\\ 
                       & 7b - 0.2 WN                                         & 0.58         & 0.77        & 0.58        & 0.58\\ 
                       & 7b - 0.3 WN                                         & 0.59           & 0.77        & 0.59          & 0.5\\ 
                       & 7b - 0.4 WN                                         & 0.59         & 0.77        & 0.59        & 0.5\\ 
                       & 7b - 0.5 WN                                         & 0.57        & 0.77        & 0.58        & 0.48\\ \hline
\end{tabular}
}
\label{table:denoising}
\end{table}

\subsection{Countermeasure: Adversarial Training} 
Here we examine the effectiveness of countermeasures against evasion attacks. The defensive techniques employed rely on adversarial training, where some adversarial examples are added to the training sets. 
The retrained models are tested against test sets explained in Experiments 3a and 3b, where 100\% of their positive samples disturbed by background noise. 
Experiments 4a and 4b examine adversarial training using samples with background noise. We use the baseline glass break and gunshot training sets from Experiment 1a, and modify them by infusing background noise to 100\% of samples in their positive class. 
Similarly, Experiments 4c and 4d take and modify the baseline 1a train sets, however ten out of eleven white noise levels (from 0.0005 to 0.5) are added proportionally to the train sets, each level, thus, perturbing two hundred samples. The retrained models are tested against the same eleven white noise infused test sets seen at Experiments 3c and 3d.

Table~\ref{table:adv-training} shows the results for these experiments. For adversarial training using sample with background noise, we achieve up to 8\%, and 29\% improvement for gunshot and and glass break, respectively. For adversarial training based using samples with white noise, we achieve nearly 50\% improvement for both gunshot and glass break.  

\subsection{Countermeasure: Denoising}
Finally, as our final defense mechanism, we try to denoise the adversarial test sets through our custom denoising function. Experiments 6a and 6b involve denoising the 100\% background noise infused test sets from Experiments 3a and 3b, while Experiment 7a and 7b involve denoising the ten white noise infused test sets from experiments 3c and 3d. The train sets are the baseline ones from Experiment 1a. 
As it can be seen in Table~\ref{table:denoising}, 7a achieves nearly 3\% accuracy improvement for both background noise denoised gunshot and glass break, while 7b achieves over 7\% improvement for white noise denoised gunshot. Experiment 10a also achieves up to a low 1\% improvement for glass break, however, given the previous unusual behavior by the glass break class, this was expected at this point. 
Despite the modest improvements in the results, it shows the potential of developing more advanced denoising techniques.

\section{Limitations}

In this work, we performed field testing focused entirely on Nest devices. We are aware other similar devices exist, and these devices may perform different. Therefore, next we test these devices. 

Another limitation of this research is its declared  focus on the home security domain. 
We consider a single choice for domain to be a limitation, as we cannot responsibly state that our results fully generalize. 
Besides, an additional domain with a established and consolidated AED application (e.g., provided by a company or organization) could allow for further specialized testing and validation of the obtained results.

\section{Future Work}

There are several research paths to be taken as next steps. We list the mainstream ones next.

    \begin{itemize}
        \item \textbf{Improvement of standalone denoising}: our denoising function offers some improvement to the performance of classifiers against adversarial examples, however we further attempt to improve the algorithm in isolation from other components of this implementation.  
      \item \textbf{Optimization of adversarial training and denoising duet}: another way to improve our denoising function may be do it in a linked fashion to adversarial training. More specifically, we might investigate if there is any correlation between improving model robustness through adversarial training and how it later reacts to denoised samples fed during testing. In other words, does adversarial training done in a specific way improve a deep learning model to the point that it, depending on the training received, it reacts differently (better or worse) to denoised samples? 
      
      \item 
      \textbf{Examining the impact of various types of noise on several types of classifiers}: Our results showed that gunshot detection classifier is more vulnerable to adversarial examples based on white noise when compared to glass breaking. One future research direction can be generating adversarial examples using other types of noise and testing them on a variety of classifiers.
    \end{itemize}

\section{Conclusions}

Our main contributions are two-fold: first, our results confirm that AED systems are vulnerable to evasion attacks by adversarial examples made of audio samples. We tested AED-capable CNNs as well as third-party devices, and while their initial baseline performance was good under ideal circumstances with regards to audio event detection, we also witnessed significant drops in classification performance, when  either background noise or white noise were injected into the audio samples. However, we noticed that not all types of noise is effective in decreasing the performance of classifiers. For example, while white noise infused to gunshot samples can significantly decrease the performance of gunshot detection classifier, while adding white noise to glass break samples do not show such effect.  
Second, we showed that defenses against these adversarial examples perform well. For instance, employing adversarial training leads to significant improvements. We also verified the potential of spectral gating denoising techniques, which when applied to our test sets, led to  better classification performance. 
As previously based, our research is done under the motivation of being one step ahead a future where Audio Event Detection Systems are going to become ubiquituous. We thus seek to motivate fellow researchers from the academy and professionals from the industry to think of potential security shortfalls before executing the design and the implementation of AED solutions, thus paving the way for a safer and more effective future.

\bibliographystyle{plain}
\bibliography{refs}

\vspace{12pt}

\end{document}